\documentclass{emulateapj}

\def\lea{\mathrel{<\kern-1.0em\lower0.9ex\hbox{$\sim$}}}
\def\gea{\mathrel{>\kern-1.0em\lower0.9ex\hbox{$\sim$}}}

\slugcomment{}

\shorttitle{Andromeda V dSph}
\shortauthors{Ata Sarajedini and Conor L. Mancone}

\begin{document}


\title{Stellar Populations in the Andromeda V Dwarf Spheroidal Galaxy\footnotemark[1]}

\author{Conor Mancone and Ata Sarajedini}
\affil{Department of Astronomy, University of Florida, Gainesville, FL 32611}
\email{cmancone@astro.ufl.edu, ata@astro.ufl.edu}



\begin{abstract}
Using archival imaging from the Wide Field Planetary Camera 2 aboard the Hubble
Space Telescope, we investigate the stellar populations of the Local Group dwarf 
spheroidal Andromeda V -  a companion satellite galaxy of M31.  The color-magnitude
diagram (CMD) extends from above the first ascent red giant branch (RGB) tip to 
approximately one magnitude below the horizontal branch (HB). The steep well-defined
RGB is indicative of a metal-poor system while the HB is populated predominantly 
redward of the RR Lyrae instability strip. Utilizing Galactic
globular cluster fiducial sequences as a reference, we calculate a mean metallicity
of $[Fe/H] = -2.20 \pm 0.15$ and a distance of $(m-M)_0 = 24.57 \pm 0.04$ 
after adopting a reddening of $E(B-V) = 0.16$. This metal abundance places And V
squarely in the absolute magnitude - metallicity diagram for dwarf spheroidal
galaxies. In addition, if we attribute the 
entire error-corrected color spread of the RGB stars to an abundance spread, we
estimate a range of $\sim$0.5 dex in the metallicities of And V stars.
Our analysis of the variable star population of And V reveals the presence of 28 potential
variables.  Of these, at least 10 are almost certainly RR Lyrae stars based on their
time sequence photometry.
%% Translating the periods of the RRab stars to metallicity, we
%% arrive at a mean abundance of $[Fe/H] = -2.07 \pm 0.30$, in good agreement with
%% the value derived from the color of the red giant branch stars. The distance determined
%% from the RR Lyrae variables of $(m-M)_0 = 24.50 \pm 0.16$ is also in excellent
%% accord with the distance calculated from our fiducial sequence analysis.
\footnotetext[1]{Based on observations with the NASA/ESA Hubble Space Telescope, obtained at the Space Telescope Science Institute}
\end{abstract}


%% subject header.  Objects should be in the appropriate "individual"
%% headers (e.g. quasars: individual, stars: individual, etc.) with the
%% additional provision that the total number of headers, including each
%% individual object, not exceed six.  The \objectname{} macro, and its
%% alias \object{}, is used to mark each object.  The macro takes the object
%% name as its primary argument.  This name will appear in the paper
%% and serve as the link's anchor in the electronic edition if the name
%% is recognized by the data centers.  The macro also takes an optional
%% argument in parentheses in cases where the data center identification
%% differs from what is to be printed in the paper.


\section{Introduction}

Dwarf spheroidal and dwarf irregular galaxies are thought to be instrumental
in the process that forms
larger galaxies (Font et al. 2006, and references therein). As such, their importance 
is sometimes considered only within
this context - that of much more massive systems. However, it is important to keep in
mind that dwarf galaxies (DGs) are useful probes of galaxy formation and evolution in and of
themselves. In this regard, the relation between the absolute magnitude of a DG and
its mean metallicity has provided a number of useful insights. First studied decades 
ago (Tinsley 1978; Mould, Kristian, \& Da Costa 1983), this relation shows that more luminous DGs possess a more metal-rich mean abundance.
This in turn suggests that self-enrichment by heavy elements is more likely in a system
with a larger gravitational potential which can retain supernova ejecta 
(e.g. Davidge et al. 2002). 

Within the context of this mass-metallicity correlation, Andromeda V, a dwarf spheroidal
companion galaxy to M31, is somewhat of an anomaly. In their discovery paper,
Armandroff et al. (1998) used Milky Way globular cluster fiducials combined with V and I imaging 
to measure a mean metallicity for And V of $[Fe/H] = -1.5$.  At its measured absolute magnitude
of $M_V \sim -9$, we expect And V to have $[Fe/H] \sim -2$, similar to the Milky Way
dSphs Ursa Minor and Draco, so the Armandroff et al. (1998)
value made And V a significant outlier among the Local Group dwarf spheroidal
galaxies in the relation between absolute magnitude and metallicity. This led Caldwell (1999)
to hypothesize that perhaps And V has a deeper than normal potential well, which could be 
verified through measurements of its stellar velocity dispersion or mass-to-light ratio.

Motivated by the discrepancy between the expected and observed properties of And V,
Davidge et al. (2002) re-examined the question of And V's chemical composition. They used 
the Gemini Multi-Object Spectrograph (GMOS) in imaging mode on the Gemini North telescope
to construct an optical color-magnitude diagram (CMD) in the g', r', and i' filters. The slope
of the And V red giant branch (RGB) was then used to calculate the mean metallicity of
[Fe/H] = $-2.2 \pm 0.1$ placing And V squarely on the $M_V - [Fe/H]$
relation.  Davidge et al. (2002) therefore asserted that there was nothing unusual about the 
metallicity of And V and that the absolute integrated magnitude and metallicity do indeed follow 
the relation for dwarf spheroidal galaxies.  
%They also noted that the mass-to-light ratio of dwarf systems was very similar among galaxies of a 
%given Mv.  
However, this result was again thrown into doubt with the work of McConnachie et al. (2005).  Using 
Johnson V and Gunn i photometry from the Isaac Newton Telescope Wide Field Camera they 
calculated $[Fe/H] = -1.6$ from the mean color of the And V RGB.  This once again returned 
And V  to the status of an outlier in the $M_V - [Fe/H]$ relation.

In light of the uncertainty about the mean abundance of stars in And V, and the overall
importance of characterizing the properties of stars in dwarf spheroidal galaxies, we have
undertaken a photometric study of And V using archival imaging from the Hubble
Space Telescope (HST) Wide FIeld Planetary Camera 2 (WFPC2). In section 2 we detail our 
observations and 
reductions, while section 3 presents the color-magnitude diagram (CMD) of And V along
with a discussion of its properties; our conclusions are summarized in Sec. 4.

\section{Observations and Reductions}

The images used in the present study are archival WFPC2 observations of And V as detailed in Table 1.  
The observations, taken as part of program GO-8272 (PI: Armandroff) 
consist of 16 F450W and 8 F555W images.  Due to a malfunction in the observing
sequence, the images were taken at two different times -- half were taken in November of 1999 and 
half in December of 2000. The first and second set of 12 observations each cover a time baseline
of approximately half of a day. There is a slight rotational offset between the two observations as 
illustrated by Fig. 1, which shows a $12$ x $12$ arcmin digitized sky survey image of And V with the  
WFPC2 footprint outlined.

All of the program frames were reduced using the HSTphot software (Dolphin 2000) package.  
First, all of
the required preliminary steps were performed such as masking the cosmic rays and the hot pixels,
as well as calculating the background sky contribution to each pixel. Then, the stellar photometry 
was performed on the processed images.  HSTphot detects stars and fits TINY TIM point spread 
functions to the detected profiles.  It also applies geometric, charge transfer efficiency, and 
aperture corrections to the resultant magnitudes. The instrumental photometry is then calibrated to 
standard Johnson-Cousins BV magnitudes (for details see Dolphin 2000). It is important to
note that the validity and robustness of Dolphin's (2000) photometry software 
and his characterization of WFPC2's photometric performance have been verified by 
Sirianni et al. (2005). 

HSTPhot is able to photometer multiple images simultaneously as long as they share the same
rotational orientation. It calculates positional offsets between the frames, matches the stars and outputs
average magnitudes in each filter. We applied this process to the two sets of images (one set at each
rotation) and found small offsets between them ($\Delta$B = 0.028, $\Delta$V =  0.010).
We then offset each set of data to the mean photometric zeropoint and combined measurements
of stars in common between the two rotations. The final
list contains stars that were detected on at least 12 different frames between the two rotations. 

% multiple images at one time.  It will take all frames given to it, calculate positional offsets, 
%calculate photmetric offsets, match the stars, 
%and then output star positions, average magnitude in each filter, and corrected magnitudes in 
%each frame.  For this to work, however, 
%all images must have the same pointing and rotation.  To get around this limitation we reduced 
%the images for each rotation separately 
%and combined the two master lists created by HSTphot.  This was done by converting star positions 
%from pixels to RA and Dec and 
%calculating the offset between the two rotations.  This offset was found to be gratifyingly small 
%(RA = $0.52 \pm .045$ arcsecs, DEC = 
%$0.43 \pm .064$ arcsecs), and was then corrected to bring the two rotations to the same WCS.  
%Stars were matched up between the 
%rotations using twice the standard deviation of the positional offset as the matching radius 
%(0.13 arcsecs).  Next a final list of stars was 
%created.  In this list all stars were included that were detected in at least 12 different frames 
%between the two rotations.  The 
%photometric offset was calculated between the two rotations ($B1-B2 = -0.028, V1-V2 = -0.010$), 
%applied, and then information from 
%all frames was combined creating a master star list of 2552 stars.

\section{Results}

\subsection{The Color Magnitude Diagram}

The CMD of And V, shown in Fig. 2, extends from above the first ascent RGB 
tip to roughly 1 magnitude below the horizontal branch (HB).  The relatively steep RGB argues 
in favor of And V being fairly metal-poor, while at the same time, the predominantly red HB
suggests a somewhat younger age (Stetson et al. 1999; Da Costa et al. 2002). 
We will explore these issues in more detail below. 
There is evidence for foreground contamination from the Milky Way on both sides of the RGB. 
Simulations using the Besancon Galaxy model (Robin et al. 2003) confirm that these are indeed 
Milky Way foreground stars as illustrated in Fig. 3 wherein the open triangles represent
the model foreground stars.  These simulations also suggest the absence of a significant intermediate-age
(2 to 8 Gyr) population in And V due to the lack of asymptotic giant branch (AGB) stars located 
above the RGB tip (Martinez-Delgado \& Aparicio 1997; Martinez-Delgado et al. 1999). 
That is to say, the number of such AGB stars is consistent with the degree 
of field contamination. This finding is in-line with that of Davidge et al. (2002) who reached 
the same conclusion based on their ground-based GMOS CMD.

\subsection{Metallicity from the Red Giant Branch}

%The upper RGB ($V < 25$) of And V shows a 1-$\sigma$ color spread around
%a mean line of  $\sim$0.1 mag in B--V.  The 
%average error in B--V in this magnitude range as given by HSTPHOT is only $\sim$0.04 mag. 
%(0.024 mags in V, 0.034 mags in B, added in quadrature).  
A statistically significant spread in the colors of And V RGB stars would suggest
the presence of a range of ages and/or metallicities in this system.  Given the lack of AGB 
stars above the first ascent tip and the relative
insensitivity of RGB colors to age as compared with metallicity, it is more likely that a
range in the colors of RGB stars is a reflection of a spread in metal abundance among
the stars in And V. We can quantify this dispersion along with the mean [Fe/H] by using
Milky Way globular cluster RGB fiducials.  For this, we make use of the sequences for
M15, NGC 6752, NGC 1851, and 47 Tucanae published by Sarajedini \& Layden (1997) and
plotted in Fig. 3. These clusters have metal abundances of --2.17, --1.54, --1.29, and --0.71, 
respectively, on the Zinn \& West (1984) scale. We have adopted the Chaboyer (1999) 
relation between HB magnitude
and metallicity - $M_V(HB) = 0.23[Fe/H] + 0.93$.

The next step is to produce a locus of points to represent the mean RGB of And V. To
expedite this, we divide the RGB stars brighter than $V=25$ into bins of 0.2 mag.
%by selecting out the RGB from the CMD (to exclude most field star contamination) and 
%then dividing the RGB stars into bins of size 0.2 mags starting at $V = 25$.  
For each bin, we used a 2-$\sigma$ rejection algorithm to calculate the mean B--V color.  
The resultant RGB locus is used to calculate the mean abundance of And V via the
procedure described by Da Costa et al. (2000). This method uses the relationship between 
metal abundance and mean HB magnitude to calculate a distance modulus based on an 
initial guess of the metal abundance; the distance is used to place the fiducials in the CMD 
and measure the mean metallicity, which is again used to calculate a new distance.  
This is an iterative process, but quickly converges after only a few calculations. To determine
the mean HB mag of And V,  we also follow the lead of Da Costa et al. (2000) by selecting
stars between $25.25 < V < 25.65$ and $-0.05 < B-V < 0.4$, which gives 
$\langle$V(HB)$\rangle$$ = 25.49 \pm 0.01$ (standard error of the mean). Adopting a
reddening of E(B--V)= 0.16  (Burstein \& Heiles, 1982), this process yields
a distance of  $(m-M)_0 = 24.57 \pm 0.04$ and a mean metallicity of $[Fe/H] = -2.20 \pm 0.15$
for And V. The errors are calculated by shifting the HB by its error and the RGB 
by the mean color error in bins of 0.2 mag and then redoing the analysis. The resultant
differences in the distance modulus and the metallicity are then the adopted
errors in these quantities.

%We applied this method to our chosen mean RGB points, and used the age corrected 
%globular cluster fiducials.  For the HB  For our initial guess of [Fe/H] we used both 
%--1.5 and --2.0, but the result is independent of the starting metallicity over the entire 
%range of reasonable abundances ([Fe/H] = --1.0 to --2.4).  Finally we calculate a distance 
%modulus of $(m-M)_0 = 24.48$ and a metallicity of $[Fe/H] = -1.96$, consistent with our 
%estimate using the original distance modulus to AndV.
%This method and the resultant mean abundance
% assume that the mean age of the stars in And V is similar to those of the Galactic
% globular clusters. We will provide a fuller discussion of this point later in the section.

%Adopting
%a distance modulus of $24.55 \pm 0.12$ (Armandroff et al. 1998) and E(B--V) of 0.16 
%(Burstein \& Heiles, 1982) allows us to compare the resultant And V RGB locus with
 %those of the Galactic globulars. An interpolation among these fiducials yields a mean
% metal abundance of [Fe/H] of $-2.12 \pm 0.16$, where the quoted error also includes the
%uncertainty in the distance modulus. 
  %As a check of our result we applied the methodology of Da Costa et al. (2000) to our data
%as well.  

We will quantify the HB morphology in the next section, but for now, it
is important to note that with such a low metallicity and a HB that is
predominantly redward of the RR Lyrae instability strip, And V represents an extreme
case of the second parameter effect. This has been noted by Harbeck et al. (2001) in their study
of this galaxy which makes use of the same observational material we present here. If we
assume that the red HB morphology of And V is primarily a result of relative youth, then
we can place some limits on how our derived mean metal abundance will change as a result.
As noted above, the lack of a significant supra-RGB-tip AGB population suggests that
the dominant population in And V is older than $\sim$8 Gyr. Yet, based on the ages and HB
morphologies of the Galactic globulars, the oldest of which have ages of around 13 Gyr, 
And V must be $\sim$3 Gyr younger than globular clusters at its metal abundance. Hence,
we estmate an age between 8 and 10 Gyr for the majority of stars in And V. This would
make our quoted mean abundance value of $[Fe/H] = -2.20 \pm 0.15$ more metal-rich
by $\sim$0.1 dex according to the theoretical models of Dotter et al. (2007).

%that this is really a lower limit on the mean metallicity of AndV.  This very metal poor result 
%combined with a predominantly red HB imply that AndV is likely much younger than the 
%typical Milky Way globular cluster.  Although the lack of a main sequence turn off makes 
%it difficult to accurately determine the age of AndV, this uncertainty in age creates a relatively 
%small error in metallicity.  To investigate the error introduced by AndV's ambiguous age, 
%we adjusted the RGB fiducials to represent a younger population.  We accomplished this 
%with the aid of the newest set of isochrones from Dotter et al. (2007).  We used the code 
%included with their isochrones to generate a set of isochrones at the metallicity of each 
%fiducial.  Then we selected the isochrones for 12Gyrs and 7Gyrs, measured the difference 
%in color between them, and shifted the fiducials by this amount.  We then recalculated the 
%metallicity of AndV using these corrected fiducials, and found that it changed to 
%$-2.01 \pm 0.18$.  We adopt this as the mean metallicity for AndV and adopt the error 
%due to the uncertain age as 0.11 dex.  Adding the error from age and distance modulus 
%in quadrature, we calculate a final mean metallicity of $[Fe/H] = -2.01 \pm 0.21$.

Moving on to the metallicity spread in And V, Fig. 5 shows the color spread of And V
stars around our adopted fiducial sequence (solid line in Fig. 4) for stars in the range
$23<V<24$. A gaussian fit to these data yields a 1-$\sigma$ color spread of 0.076 mag, which
becomes 0.070 mag after the mean photometric error of 0.031 mag is subtracted
in quadrature.  We note that this mean error as given by HSTphot is consistent with
the dispersion in the magnitudes of the same stars measured at the two rotations.
Using the globular cluster fiducials plotted in Fig. 4, we can use the 
dereddened color of these sequences at $M_V = -1.5$, the approximate middle of
the magnitude range of RGB stars we are considering, as a function of abundance to
translate a color range of 0.070 mag to a metal abundance spread of 0.52 dex.

Converting the intrinsic color spread to a metallicity spread is complicated
by two effects. First, there is the possibility that some of the color spread is caused
by a range of ages among the And V stellar population. However, as discussed above,
based on the lack of a young main sequence and AGB stars above the first ascent RGB
tip, and the relative insensitivity of RGB colors to age, we expect this effect to be
be small. Second, there is the problem of needing to extrapolate the globular
cluster fiducials to more metal-poor regimes in order to convert the RGB color spread
to an abundance spread. Both of these effects will introduce a level of uncertainty
into our metallicity spread determination.

\subsection{Horizontal Branch Morphology}

It is clear from the CMD of And V that this galaxy has a primarily red HB, one explanation 
of which is the presence of a substantial young population.  To quantify the HB morphology 
of And V, we follow the procedure outlined in Da Costa et al. (2000) for calculating the 
index $i = b/(b + r)$, where b and r represent the number of stars on the blue and red sides
of the RR Lyrae instability strip, respectively.  This method was originally developed for 
their work on And I and And II (Da Costa et al. 1996, Da Costa et al. 2000).  In the case 
of those two companion galaxies, one set of HB color and magnitude limits was used.  
However the lower metallicity of And III and And V shifts their RGBs to the blue, 
introducing RGB contamination into the HB if the same color limits are used.  To circumvent 
this difficulty for And III, the HB morphology index was measured 
by Da Costa et al. (2002) using two different methods - a procedure that we also adopt here.

For the first calculation, a color histogram is created for the stars on the HB 
($25.25 < F555W < 25.65$).  This histogram exhibits a sharp decline redward of 
approximately $F450W - F555W = 0.5$, as was the case for Andromeda III.  We take 
this to be the point at which RGB stars begin to dominate over the HB stars.  We then 
take the red stars to be those between these magnitude limits and between the colors 
$0.35 < F450W - F555W < 0.5$, and find that $r = 339$.  Next we use the same magnitude 
range for the blue stars and use color limits of $-0.05 < F450W - F555W < 0.25$, finding 
$b = 107$.  Assuming errors due only to Poisson statistics, we find $i' = 0.24 \pm 0.03$. 
The prime designates a value based on the And III color limits.
% where the subscript "iii" denotes that we have used the same color limits that 
%Da Costa used for Andromeda III.

Our next approach involves using the same color limits as used with And I and And II -
the red edge of the HB is placed at $F450W - F555W = 0.60$ - and then correcting for 
RGB contamination by using the RGB density above and below the HB to estimate 
the number of contaminating stars.  This yields $r = 416$, 
$b = 107$, and $i = 0.20 \pm 0.02$, in agreement with our initial estimate.  Compared with 
values of $i=0.13 \pm 0.01$, $0.18 \pm 0.02$, and $0.10 \pm 0.02$ for And I, And II, and 
And III, respectively, And V is on the blue end of the HB morphology scale for these dwarf 
galaxies, though it is similar to And III.

Another important point of investigation is the presence of a radial gradient in the HB 
morphology of And V. Previous searches for this effect have revealed a gradient in And I, 
but not in And II or And III.  We have searched for a gradient in And V using the same 
methodology as Da Costa et al. (2000).  
First, adopting a core radius of 27.91" and zero eccentricity (Caldwell 1999),  we divide 
the stars into two populations: those inside the core radius and those outside. 
We find $i' = 0.20 \pm 0.07$ and $i = 0.11 \pm 0.03$ for the inner population 
and $i' = 0.24 \pm 0.03$ and $i = 0.23 \pm 0.03$ for the outer population.  Next we divide
the stars in And V at a distance of 52" to yield two equal-sized samples.  For these 
two populations we find $i' = 0.24 \pm 0.04$ and $i = 0.17 \pm 0.03$ for the inner 
stars and $i' = 0.24 \pm 0.03$ and $i = 0.24 \pm 0.03$ 
for the outer stars.  The primary conclusion from this exercise is that if there is a radial gradient in the 
HB morphology of And V, it is too small to be reliably detected.  This lack of a population gradient is in 
agreement with the results of Harbeck et al. (2001).

\subsection{Characterization of Variable Stars}

To identify the variable stars in And V we created an initial candidate list for each rotation separately.  To be 
considered as a candidate, a star had to be detected in every exposure of a given rotation and exhibit a 
frame-to-frame standard deviation of $\geq$0.2 mag.  We then examined the raw light curves of the 
resulting candidates from each rotation to identify potential RR Lyrae stars.  This process provided 
a total of 28 RR Lyrae candidates.  When available, the photometric data from the other rotation was 
then added to the dataset and used as input into our period finding routine.  Ten candidates had data 
from both observing windows.  These ten variables have $<$V$>$ = 25.55 and $<$B-V$>$ = 0.30 putting them in the middle 
of the HB, which combined with the shapes of their raw light curves unambiguously identifies them as RR Lyraes.
We therefore refer to these 10 RR Lyraes as our high confidence variables.  The other 18 candidates are also 
likely to be RR Lyraes but due to poor phase coverage we can't exclude the possibility that they are another 
type of variable.  These form our set of candidate variables.  Tables 2-7 show the raw magnitude 
measurements at each epoch for all variables. Table 3 lists RA and Dec for the high confidence variables.  The locations of 
the high confidence variables in the CMD are shown in Fig. 3.  Table 4 lists positions for the candidate variables.

We attempted to measure periods for our high confidence variables by using a template fitting method similar to that
of Layden (1998).  We iterated with a step size of 0.0001 days over the range of periods from 0.2-1.5 days, and at each 
point used Pikaia (Charbonneau 1995) to find the combination of epoch, amplitude, and mean magnitude that minimized 
the $\chi$$^2$ differences between the observed data points and 10 variable star templates taken from the work of 
Layden (1998).  Pikaia was then run once more with period as an additional free parameter, allowing it to search within 
$\pm$0.0001 days of the best fitting period from the inital search.  This refinement step provided our best guess
for the period of each variable.  As a check of our results a series of statistical simulations were 
performed in which artificial RR Lyraes were created with the same photometric 
errors and observing cadence as the observations, and then fitted in the same manner as our high confidence variables.  
These simulations demonstrated that our template fitting algorithm couldn't robustly derive periods for the variables.  
Unfortunately, this result is not a limit of our algorithm but a result of the spacing of the observations.  The 
longest contiguous observing window in a single filter is only 0.4 days, too short to accurately measure the likely 
periods of the AB-type RR Lyraes (0.5-0.8 days).  Although we cannot measure periods for these variables we include 
figure 6, a plot of the raw and folded light curves for high confidence variable 8, to demonstrate the RR Lyrae 
nature of these variables.  The parameters of the fit come from our template fitting algorithm.  The fitted 
parameters for this variable are $<$B$>$ = 25.87, $<$V$>$ = 25.31, B amplitude = 1.03, V amplitude = 0.83, a 
period of 0.785235 days, and a starting epoch of 2451496.04345703 (HJD).  This variable demonstrates the typical 
quality of the raw and fitted light curves for the high confidence variables.

\section{Summary and Conclusions}

This work presents HST WFPC2 F450W and F555W observations of Andromeda V, a dwarf spheroidal
galaxy satellite of M31.  Comparing the CMD of And V to Milky Way Globular cluster fiducials of known 
metallicity, we find a mean metal abundance for And V of $[Fe/H] = -2.20 \pm 0.15$ with a
spread of $\sim$0.5 dex, and a distance 
modulus of $(m-M)_0 = 24.57 \pm 0.04$.  This result puts Andromeda V 
squarely on the absolute magnitude - metallicity relationship for local group dwarf spheroidals.  As 
suggested by Davidage et al. (2004) this tightens the relationship between absolute magnitude and 
metallicity, suggesting that there is little scatter in the relationship between mass-to-light ratio and 
absolute magnitude for dwarf spheroidals.  In addition we find RR Lyraes in And V providing evidence for an old population,
but are unable to accurately measure their periods.

\acknowledgments

We are grateful to Andy Layden for providing his suite of software for light curve fitting and providing
useful input in the process of applying and modifying the software. We are also grateful to Michael Barker 
for providing suggestions for the paper and recommending pikaia as an alternate optimization algorithm.  The comments of an anonymous referee greatly clarified the presentation
in the manuscript.  This research was supported by grant number AR-11277.01-A provided by NASA
through the Space Telescope Science Institute, which is operated by the
Association of Universities for Research in Astronomy, Incorporated, under
NASA contract NAS5-26555.

\clearpage

%table 1
\begin{deluxetable*}{lcccccc}
\tablecaption{Observing Log}
\tablewidth{0pt}
\tablehead{
  \colhead{Date}
  &\colhead{Dataset}
  &\colhead{Filter}
  &\colhead{Exp Time}
}
\startdata
November 11, 1999 & U5C701 &F555W & 3 x 1200s\\
November 12, 1999 & U5C701 & F450W & 8 x 1200s\\
November 13, 1999 & U5C702 & F555W & 1 x 1300s\\
December 16, 2000 & U5C752 & F555W & 1 x 1200s\\
December 17, 2000 & U5C752 & F555W & 3 x 1200s\\
December 17, 2000 & U5C752 & F450W & 8 x 1300s
\enddata
\end{deluxetable*}

%table 2a
\begin{deluxetable*}{cccccccccccc}
\tablecaption{Observed Light Curves For Variables 01-05}
\tabletypesize{\scriptsize}
\tablewidth{0pt}
\tablehead{
  \colhead{Filter}
  &\colhead{HJD}
  &\colhead{01 Mag} &\colhead{01 Err}
  &\colhead{02 Mag} &\colhead{02 Err}
  &\colhead{03 Mag} &\colhead{03 Err}
  &\colhead{04 Mag} &\colhead{04 Err}
  &\colhead{05 Mag} &\colhead{05 Err}
}
\startdata
V &  2451494.37626  &  25.42  &  0.093  &  25.33  &  0.083  &  25.69  &  0.111  &  25.54  &  0.100  &  25.23  &  0.076\\
V &  2451494.39432  &  25.61  &  0.114  &  25.15  &  0.077  &  25.83  &  0.129  &  25.71  &  0.118  &  25.16  &  0.079\\
V &  2451494.44154  &  25.61  &  0.110  &  25.63  &  0.105  &  25.64  &  0.106  &  25.36  &  0.086  &  25.36  &  0.085\\
V &  2451494.50821  &  25.88  &  0.135  &  25.71  &  0.112  & \nodata & \nodata &  25.37  &  0.087  &  25.43  &  0.114\\
B &  2451494.57614  &  26.51  &  0.255  & \nodata & \nodata & \nodata & \nodata &  25.55  &  0.118  &  26.28  &  0.173\\
B &  2451494.64350  &  26.72  &  0.306  &  25.45  &  0.108  &  25.55  &  0.114  &  25.71  &  0.135  &  26.08  &  0.148\\
B &  2451494.71017  &  26.48  &  0.253  &  25.52  &  0.115  &  25.73  &  0.130  &  25.66  &  0.128  &  26.20  &  0.189\\
B &  2451494.77753  &  25.87  &  0.149  &  25.83  &  0.143  &  26.15  &  0.177  &  25.84  &  0.146  &  26.52  &  0.207\\
B &  2451494.84420  &  25.48  &  0.120  &  26.12  &  0.244  &  26.34  &  0.205  &  25.44  &  0.108  &  26.46  &  0.199\\
B &  2451494.91156  &  25.55  &  0.116  &  25.56  &  0.116  &  26.09  &  0.169  &  25.24  &  0.092  &  25.66  &  0.134\\
B &  2451494.93031  &  25.84  &  0.150  &  25.56  &  0.120  &  26.50  &  0.252  &  25.43  &  0.110  &  25.43  &  0.093\\
B &  2451494.97892  &  26.17  &  0.190  &  25.50  &  0.114  &  25.90  &  0.147  &  25.68  &  0.129  &  25.95  &  0.169\\
V &  2451895.46983  &  25.64  &  0.116  &  25.56  &  0.104  &  25.38  &  0.091  &  25.28  &  0.078  &  25.76  &  0.118\\
V &  2451895.53511  &  25.65  &  0.121  &  25.17  &  0.079  &  25.50  &  0.113  &  25.62  &  0.156  &  25.62  &  0.109\\
V &  2451895.60177  &  25.53  &  0.109  &  25.44  &  0.096  &  25.89  &  0.139  &  25.66  &  0.105  &  25.69  &  0.116\\
V &  2451895.66914  &  24.91  &  0.068  &  25.84  &  0.134  &  25.55  &  0.106  &  25.82  &  0.125  &  25.54  &  0.103\\
B &  2451895.73638  &  25.42  &  0.115  &  26.08  &  0.183  &  26.81  &  0.338  &  26.49  &  0.234  &  25.39  &  0.111\\
B &  2451895.75513  &  25.49  &  0.109  &  26.03  &  0.160  &  26.31  &  0.222  &  26.31  &  0.185  &  25.28  &  0.094\\
B &  2451895.80374  &  25.65  &  0.136  &  25.75  &  0.137  &  26.18  &  0.190  &  26.09  &  0.164  &  25.36  &  0.106\\
B &  2451895.82249  &  25.74  &  0.132  &  25.50  &  0.106  &  26.19  &  0.174  &  25.82  &  0.132  &  25.57  &  0.116\\
B &  2451895.87041  &  25.73  &  0.143  &  25.54  &  0.117  &  25.86  &  0.148  &  26.04  &  0.157  &  25.84  &  0.157\\
B &  2451895.88916  &  26.03  &  0.167  &  25.96  &  0.151  &  25.91  &  0.144  &  25.94  &  0.139  &  25.64  &  0.122\\
B &  2451895.93777  &  26.13  &  0.197  &  25.94  &  0.159  &  25.76  &  0.136  &  26.21  &  0.183  &  25.96  &  0.171\\
B &  2451895.95652  &  26.13  &  0.181  &  26.11  &  0.167  &  25.54  &  0.107  &  26.28  &  0.183  &  25.94  &  0.154\\
\enddata
\end{deluxetable*}

%table 2b
\clearpage
\begin{deluxetable*}{cccccccccccc}
\tablecaption{Observed Light Curves For Variables 06-10}
\tabletypesize{\scriptsize}
\tablewidth{0pt}
\tablehead{
  \colhead{Filter}
  &\colhead{HJD}
  &\colhead{06 Mag} &\colhead{06 Err}
  &\colhead{07 Mag} &\colhead{07 Err}
  &\colhead{08 Mag} &\colhead{08 Err}
  &\colhead{09 Mag} &\colhead{09 Err}
  &\colhead{10 Mag} &\colhead{10 Err}
}
\startdata
V &  2451494.37626  &  26.16  &  0.157  &  25.48  &  0.093  &  25.71  &  0.109  &  25.37  &  0.085  &  25.65  &  0.099\\
V &  2451494.39432  &  26.10  &  0.157  &  25.58  &  0.107  &  25.68  &  0.115  &  25.32  &  0.084  &  25.67  &  0.107\\
V &  2451494.44154  &  25.61  &  0.108  &  25.55  &  0.099  &  24.95  &  0.063  &  25.55  &  0.097  &  25.69  &  0.101\\
V &  2451494.50821  &  25.36  &  0.107  & \nodata & \nodata & \nodata & \nodata &  25.60  &  0.119  &  25.76  &  0.108\\
B &  2451494.57614  &  26.15  &  0.163  &  25.11  &  0.086  &  25.84  &  0.130  &  26.19  &  0.169  &  26.17  &  0.180\\
B &  2451494.64350  &  26.55  &  0.209  &  25.39  &  0.105  &  26.07  &  0.158  &  26.37  &  0.191  &  25.92  &  0.148\\
B &  2451494.71017  & \nodata & \nodata &  25.64  &  0.126  &  25.89  &  0.137  &  26.75  &  0.260  &  25.86  &  0.140\\
B &  2451494.77753  &  26.92  &  0.286  &  25.72  &  0.133  &  26.01  &  0.151  &  26.29  &  0.187  &  25.12  &  0.081\\
B &  2451494.84420  &  26.60  &  0.219  &  25.72  &  0.134  &  26.15  &  0.168  &  25.58  &  0.104  &  25.45  &  0.103\\
B &  2451494.91156  & \nodata & \nodata & \nodata & \nodata &  26.27  &  0.188  &  25.89  &  0.130  &  25.61  &  0.116\\
B &  2451494.93031  &  26.91  &  0.289  &  26.01  &  0.176  &  25.95  &  0.152  &  25.86  &  0.131  &  25.57  &  0.116\\
B &  2451494.97892  &  26.74  &  0.258  &  25.91  &  0.153  & \nodata & \nodata &  25.87  &  0.127  &  26.13  &  0.172\\
V &  2451895.46983  &  25.78  &  0.118  &  25.54  &  0.102  &  25.44  &  0.096  &  25.57  &  0.103  &  25.57  &  0.098\\
V &  2451895.53511  &  25.95  &  0.138  &  25.57  &  0.107  &  25.42  &  0.098  &  25.39  &  0.094  &  25.61  &  0.104\\
V &  2451895.60177  &  25.82  &  0.126  &  25.45  &  0.097  &  25.61  &  0.113  &  25.41  &  0.094  &  25.34  &  0.085\\
V &  2451895.66914  &  26.03  &  0.152  &  25.52  &  0.104  &  25.29  &  0.090  &  25.16  &  0.079  & \nodata & \nodata\\
B &  2451895.73638  &  25.36  &  0.118  &  24.90  &  0.085  &  25.27  &  0.104  &  25.81  &  0.137  &  25.65  &  0.118\\
B &  2451895.75513  &  24.97  &  0.080  &  24.80  &  0.075  &  25.27  &  0.098  &  25.86  &  0.127  &  25.66  &  0.110\\
B &  2451895.80374  &  25.29  &  0.110  &  25.21  &  0.107  &  25.43  &  0.120  &  25.45  &  0.102  &  26.02  &  0.154\\
B &  2451895.82249  &  25.32  &  0.105  &  25.07  &  0.090  &  25.53  &  0.118  &  25.81  &  0.126  &  25.86  &  0.127\\
B &  2451895.87041  &  25.75  &  0.177  &  25.43  &  0.132  &  25.60  &  0.181  &  26.03  &  0.161  &  26.05  &  0.159\\
B &  2451895.88916  &  25.52  &  0.120  &  25.25  &  0.103  &  25.73  &  0.138  &  25.93  &  0.137  &  26.13  &  0.155\\
B &  2451895.93777  &  25.78  &  0.162  &  25.98  &  0.201  &  25.73  &  0.148  &  26.17  &  0.180  &  26.38  &  0.204\\
B &  2451895.95652  &  25.76  &  0.144  &  25.68  &  0.146  &  25.99  &  0.169  &  26.15  &  0.161  &  26.25  &  0.169\\
\enddata
\end{deluxetable*}

%table 2c
\begin{deluxetable*}{cccccccccccc}
\tablecaption{Observed Light Curves For Variables 11-15}
\tabletypesize{\scriptsize}
\tablewidth{0pt}
\tablehead{
  \colhead{Filter}
  &\colhead{HJD}
  &\colhead{11 Mag} &\colhead{11 Err}
  &\colhead{12 Mag} &\colhead{12 Err}
  &\colhead{13 Mag} &\colhead{13 Err}
  &\colhead{14 Mag} &\colhead{14 Err}
  &\colhead{15 Mag} &\colhead{15 Err}
}
\startdata
V &  2451895.46983  &  25.25  &  0.081  &  25.70  &  0.111  &  25.58  &  0.105  &  25.91  &  0.129  &  25.31  &  0.087\\
V &  2451895.53511  &  25.38  &  0.090  &  25.73  &  0.117  &  25.34  &  0.087  &  25.72  &  0.111  &  25.65  &  0.116\\
V &  2451895.60177  &  25.34  &  0.089  &  25.50  &  0.098  &  25.45  &  0.096  &  25.66  &  0.112  &  25.46  &  0.101\\
V &  2451895.66914  &  25.66  &  0.114  &  25.79  &  0.123  &  25.77  &  0.168  &  25.61  &  0.104  &  25.01  &  0.072\\
B &  2451895.73638  &  26.19  &  0.219  &  25.06  &  0.089  &  26.19  &  0.191  &  25.96  &  0.160  &  25.37  &  0.105\\
B &  2451895.75513  &  26.00  &  0.151  &  25.11  &  0.087  &  26.20  &  0.179  &  26.23  &  0.190  &  25.49  &  0.104\\
B &  2451895.80374  &  26.22  &  0.187  &  25.30  &  0.105  &  26.43  &  0.225  &  26.37  &  0.213  &  25.72  &  0.135\\
B &  2451895.82249  &  26.34  &  0.200  &  25.33  &  0.101  &  26.38  &  0.206  &  26.55  &  0.244  &  25.66  &  0.119\\
B &  2451895.87041  &  26.27  &  0.187  &  25.59  &  0.131  &  26.34  &  0.228  &  26.33  &  0.236  &  25.89  &  0.157\\
B &  2451895.88916  &  26.50  &  0.228  &  25.70  &  0.132  &  26.84  &  0.329  &  25.90  &  0.152  &  25.84  &  0.141\\
B &  2451895.93777  &  26.49  &  0.223  &  26.03  &  0.181  &  26.87  &  0.332  &  25.84  &  0.141  &  26.11  &  0.188\\
B &  2451895.95652  &  26.62  &  0.262  &  25.80  &  0.182  &  26.20  &  0.185  &  25.96  &  0.154  &  25.84  &  0.138\\
\enddata
\end{deluxetable*}

%table 2d
\begin{deluxetable*}{cccccccccccc}
\tablecaption{Observed Light Curves For Variables 16-20}
\tabletypesize{\scriptsize}
\tablewidth{0pt}
\tablehead{
  \colhead{Filter}
  &\colhead{HJD}
  &\colhead{16 Mag} &\colhead{16 Err}
  &\colhead{17 Mag} &\colhead{17 Err}
  &\colhead{18 Mag} &\colhead{18 Err}
  &\colhead{19 Mag} &\colhead{19 Err}
  &\colhead{20 Mag} &\colhead{20 Err}
}
\startdata
V &  2451895.46983  &  25.00  &  0.068  &  25.61  &  0.125  &  25.66  &  0.109  &  25.71  &  0.113  &  25.01  &  0.067\\
V &  2451895.53511  &  25.29  &  0.087  &  25.57  &  0.118  &  25.38  &  0.092  &  25.86  &  0.132  &  25.03  &  0.069\\
V &  2451895.60177  &  25.58  &  0.106  &  25.68  &  0.131  &  25.10  &  0.074  &  25.99  &  0.145  &  25.01  &  0.067\\
V &  2451895.66914  &  25.62  &  0.112  &  24.99  &  0.075  &  25.31  &  0.089  &  25.51  &  0.102  &  24.95  &  0.065\\
B &  2451895.73638  &  26.38  &  0.213  &  25.44  &  0.120  &  25.98  &  0.152  &  25.21  &  0.096  &  25.80  &  0.124\\
B &  2451895.75513  &  26.46  &  0.199  &  25.55  &  0.122  &  25.96  &  0.133  &  25.29  &  0.093  &  25.30  &  0.123\\
B &  2451895.80374  &  26.11  &  0.169  &  25.70  &  0.146  &  26.24  &  0.181  &  25.76  &  0.141  &  25.65  &  0.109\\
B &  2451895.82249  &  25.92  &  0.132  &  25.87  &  0.160  &  25.82  &  0.122  &  25.66  &  0.122  &  25.97  &  0.144\\
B &  2451895.87041  &  25.66  &  0.120  &  25.95  &  0.182  &  26.27  &  0.186  &  25.89  &  0.157  &  25.99  &  0.140\\
B &  2451895.88916  &  25.48  &  0.098  &  26.17  &  0.226  &  26.12  &  0.154  &  26.28  &  0.232  &  25.88  &  0.120\\
B &  2451895.93777  &  25.75  &  0.127  &  25.93  &  0.177  &  25.82  &  0.148  &  26.25  &  0.208  &  25.83  &  0.124\\
B &  2451895.95652  &  25.90  &  0.133  &  25.93  &  0.164  &  25.59  &  0.103  &  26.09  &  0.165  &  25.86  &  0.118\\
\enddata
\end{deluxetable*}

%table 2e
\begin{deluxetable*}{cccccccccccc}
\tablecaption{Observed Light Curves For Variables 21-24}
\tabletypesize{\scriptsize}
\tablewidth{0pt}
\tablehead{
  \colhead{Filter}
  &\colhead{HJD}
  &\colhead{21 Mag} &\colhead{21 Err}
  &\colhead{22 Mag} &\colhead{22 Err}
  &\colhead{23 Mag} &\colhead{23 Err}
  &\colhead{24 Mag} &\colhead{24 Err}
}
\startdata
V &  2451895.46983  &  25.56  &  0.111  &  25.73  &  0.111  &  25.84  &  0.138 & 25.66 &  0.111\\
V &  2451895.53511  &  25.05  &  0.076  &  25.36  &  0.086  &  25.61  &  0.118 & 25.94 &  0.139\\
V &  2451895.60177  &  25.21  &  0.087  &  25.11  &  0.071  &  25.66  &  0.122 & 25.73 &  0.119\\
V &  2451895.66914  &  25.36  &  0.098  &  25.07  &  0.070  &  25.10  &  0.079 & 25.68 &  0.112\\
B &  2451895.73638  &  25.83  &  0.172  &  25.77  &  0.123  &  25.28  &  0.111 & 25.56 &  0.208\\
B &  2451895.75513  &  25.77  &  0.196  &  25.84  &  0.120  &  25.47  &  0.120 & 24.93 &  0.088\\
B &  2451895.80374  &  26.25  &  0.253  &  25.77  &  0.121  &  25.47  &  0.126 & 25.12 &  0.138\\
B &  2451895.82249  &  25.86  &  0.152  &  25.91  &  0.125  &  25.75  &  0.149 & 25.15 &  0.103\\
B &  2451895.87041  &  25.91  &  0.172  &  26.12  &  0.159  &  25.98  &  0.199 & 25.48 &  0.162\\
B &  2451895.88916  &  25.84  &  0.150  &  26.10  &  0.142  &  25.87  &  0.163 & 25.58 &  0.143\\
B &  2451895.93777  &  25.49  &  0.119  &  26.36  &  0.188  &  25.87  &  0.199 & 25.95 &  0.195\\
B &  2451895.95652  &  25.39  &  0.104  &  25.77  &  0.113  &  25.84  &  0.159 & 25.57 &  0.141\\
\enddata
\end{deluxetable*}

%table 2f
\begin{deluxetable*}{cccccccccccc}
\tablecaption{Observed Light Curves For Variables 25-28}
\tabletypesize{\scriptsize}
\tablewidth{0pt}
\tablehead{
  \colhead{Filter}
  &\colhead{HJD}
  &\colhead{25 Mag} &\colhead{25 Err}
  &\colhead{26 Mag} &\colhead{26 Err}
  &\colhead{27 Mag} &\colhead{27 Err}
  &\colhead{28 Mag} &\colhead{28 Err}
}
\startdata
V &  2451494.37626  &  25.24  &  0.078  &  25.50  &  0.095  &  24.94  &  0.068  &  25.10  &  0.068\\
V &  2451494.39432  &  25.19  &  0.079  &  25.49  &  0.099  &  24.94  &  0.071  &  25.18  &  0.077\\
V &  2451494.44154  &  25.44  &  0.091  &  25.57  &  0.101  &  25.12  &  0.078  &  25.43  &  0.086\\
V &  2451494.50821  &  25.34  &  0.085  &  25.36  &  0.087  &  25.35  &  0.092  &  25.56  &  0.096\\
B &  2451494.57614  &  26.05  &  0.151  &  26.23  &  0.176  &  25.84  &  0.146  &  25.94  &  0.145\\
B &  2451494.64350  &  25.89  &  0.135  &  26.31  &  0.189  &  25.48  &  0.126  &  26.28  &  0.192\\
B &  2451494.71017  &  25.48  &  0.101  &  26.48  &  0.219  &  26.04  &  0.175  &  26.07  &  0.162\\
B &  2451494.77753  &  25.63  &  0.109  &  26.54  &  0.222  &  25.98  &  0.167  &  25.80  &  0.138\\
B &  2451494.84420  &  25.91  &  0.136  &  26.26  &  0.285  &  25.94  &  0.161  &  25.37  &  0.094\\
B &  2451494.91156  &  25.87  &  0.130  &  25.95  &  0.139  &  25.61  &  0.122  &  25.74  &  0.122\\
B &  2451494.93031  &  26.08  &  0.156  &  25.78  &  0.138  &  25.64  &  0.130  &  25.39  &  0.097\\
B &  2451494.97892  &  26.39  &  0.196  &  26.20  &  0.170  &  25.27  &  0.095  &  25.69  &  0.118\\
\enddata
\end{deluxetable*}

%table 3
\begin{deluxetable*}{cccccccccc}
\tablecaption{High Confidence RR Lyraes}
\tablewidth{0pt}
\tablehead{
  \colhead{id}
  &\colhead{RA}
  &\colhead{Dec}
}
\startdata
1  & 1:10:10.67 & 47:37:44.32\\
2  & 1:10:13.00 & 47:37:36.32\\
3  & 1:10:17.06 & 47:37:54.11\\
4  & 1:10:20.09 & 47:37:46.14\\
5  & 1:10:19.21 & 47:38:13.30\\
6  & 1:10:16.07 & 47:38:44.05\\
7  & 1:10:17.54 & 47:36:14.61\\
8  & 1:10:14.73 & 47:37:02.52\\
9  & 1:10:16.66 & 47:38:29.97\\
10 & 1:10:18.43 & 47:36:53.15\\
\enddata
\end{deluxetable*}

\clearpage

%table 4
\begin{deluxetable*}{cccc}
\tablecaption{Additional Variable Candidates}
\tablewidth{0pt}
\tablehead{
  \colhead{id}
  &\colhead{RA}
  &\colhead{Dec}
}
\startdata
11 & 1:10:20.56 & 47:37:33.76\\
12 & 1:10:15.16 & 47:37:27.14\\
13 & 1:10:19.46 & 47:37:28.65\\
14 & 1:10:19.16 & 47:37:38.38\\
15 & 1:10:18.69 & 47:38:43.66\\
16 & 1:10:18.08 & 47:38:01.63\\
17 & 1:10:20.53 & 47:39:00.35\\
18 & 1:10:17.41 & 47:38:18.93\\
19 & 1:10:17.74 & 47:38:09.37\\
20 & 1:10:19.03 & 47:37:16.91\\
21 & 1:10:13.55 & 47:36:52.50\\
22 & 1:10:18.38 & 47:37:05.46\\
23 & 1:10:16.89 & 47:36:05.58\\
24 & 1:10:22.16 & 47:37:41.89\\
25 & 1:10:17.35 & 47:37:34.27\\
26 & 1:10:14.24 & 47:37:37.51\\
27 & 1:10:23.81 & 47:38:02.53\\
28 & 1:10:15.78 & 47:37:20.85\\
\enddata
\end{deluxetable*}

%% the file, uncomment the \plotone, \plottwo, and
%% If you need a layout that cannot be achieved with \plotone or
%% \plottwo, you can invoke the graphicx package directly with the
%% \includegraphics command or use \plotfiddle. For more information,
%% please see the tutorial on "Using Electronic Art with AASTeX" in the
%% documentation section at the AASTeX Web site,
%% http://www.journals.uchicago.edu/AAS/AASTeX.
%%
%% The examples below also include sample markup for submission of
%% supplemental electronic materials. As always, be sure to check
%% the instructions to authors for the journal you are submitting to
%% for specific submissions guidelines as they vary from

%% This example uses \plotone to include an EPS file scaled to
%% 80% of its natural size with \epsscale. Its caption
%% has been written to indicate that additional figure parts will be
%% available in the electronic journal.

\clearpage
\begin{figure}
%Fig. 1
\epsscale{0.9}
\plotone{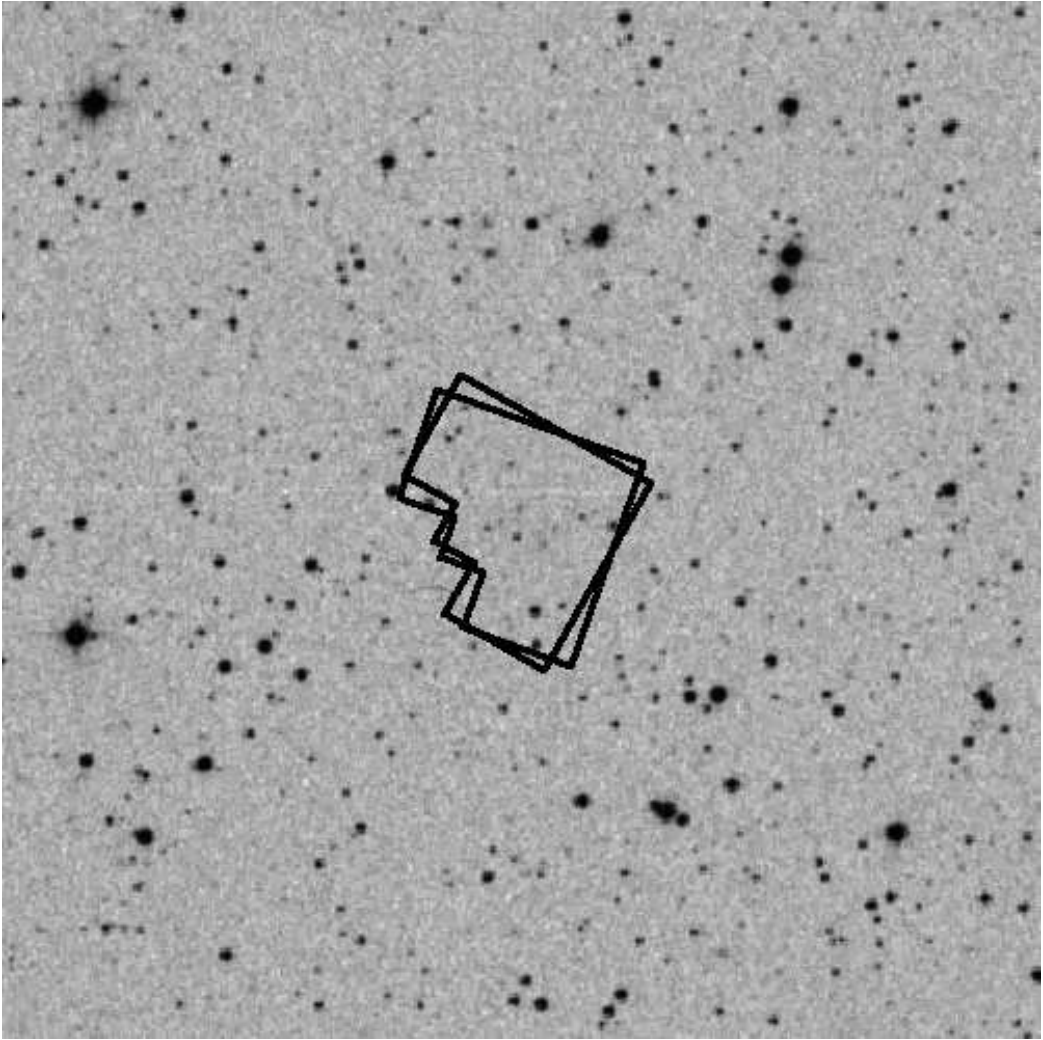}
\caption{The location of our WFPC2 fields overplotted on
a digitized sky survey image of Andromeda V. The field is approximately 12 arcmin a side;
North is up and east is to the left.}
\end{figure}

\begin{figure}
%Fig. 2
\epsscale{0.9}
\plotone{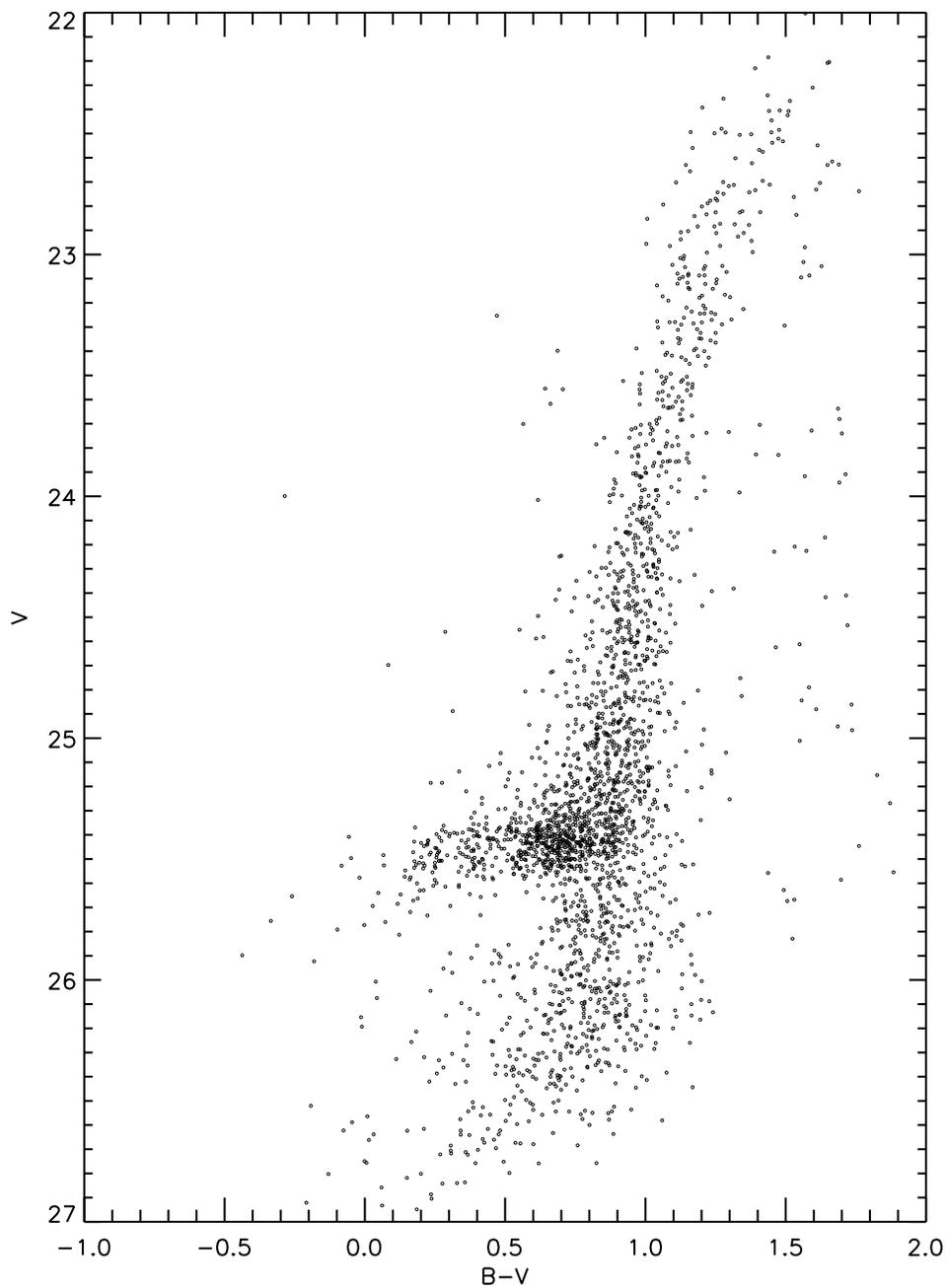}
\caption{The color-magnitude diagram of And V derived from archival HST/WFPC2
imaging.}
\end{figure}

\begin{figure}
%Fig. 3
\epsscale{0.9}
\plotone{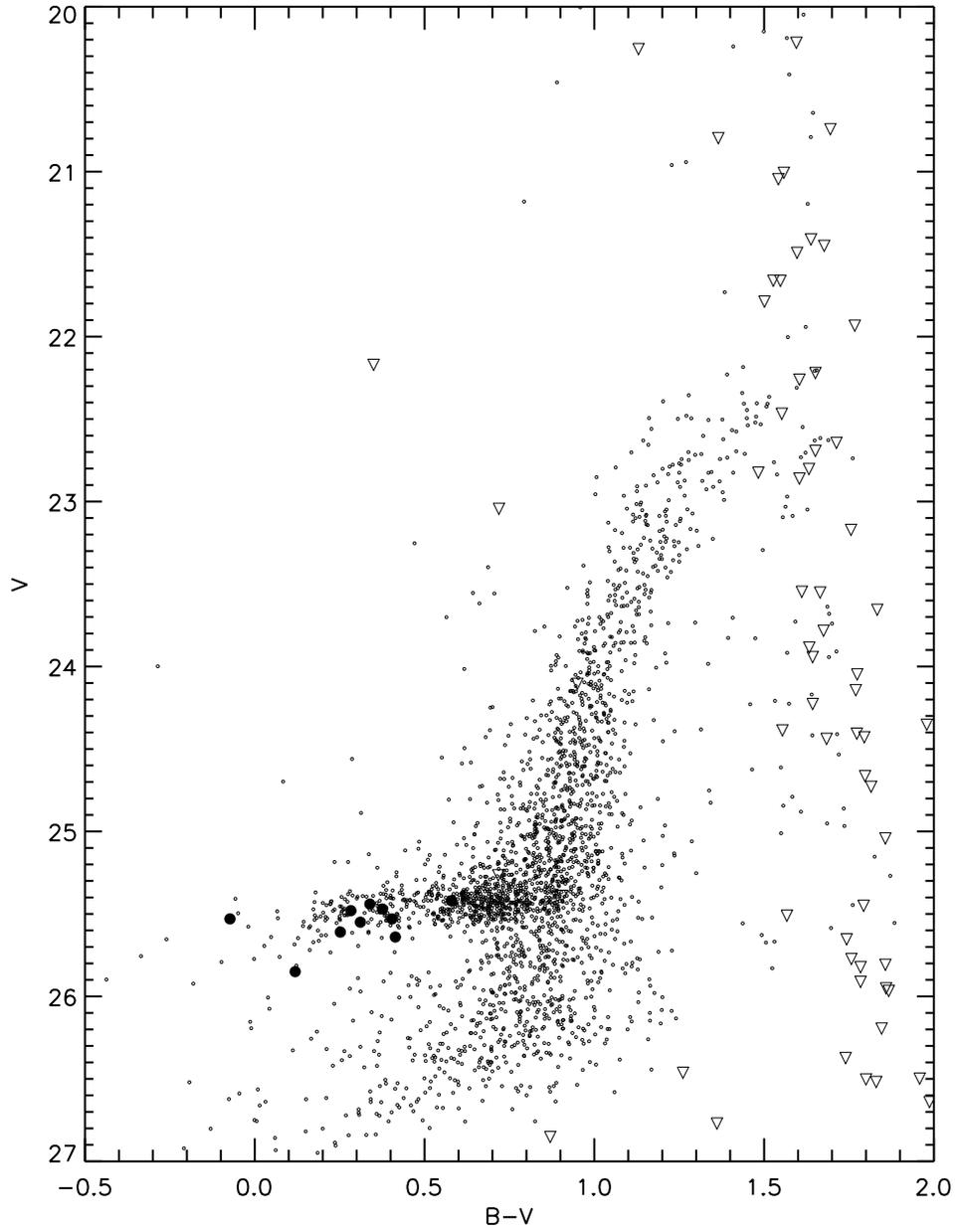}
\caption{The color-magnitude diagram of And V showing the locations of the 10 high
confidence RR Lyrae variables (filled
circles). The open triangles are the simulated field stars created using the Besancon 
Galaxy model of Robin et al. (2003).}
\end{figure}

\begin{figure}
%Fig. 4
\epsscale{0.9}
\plotone{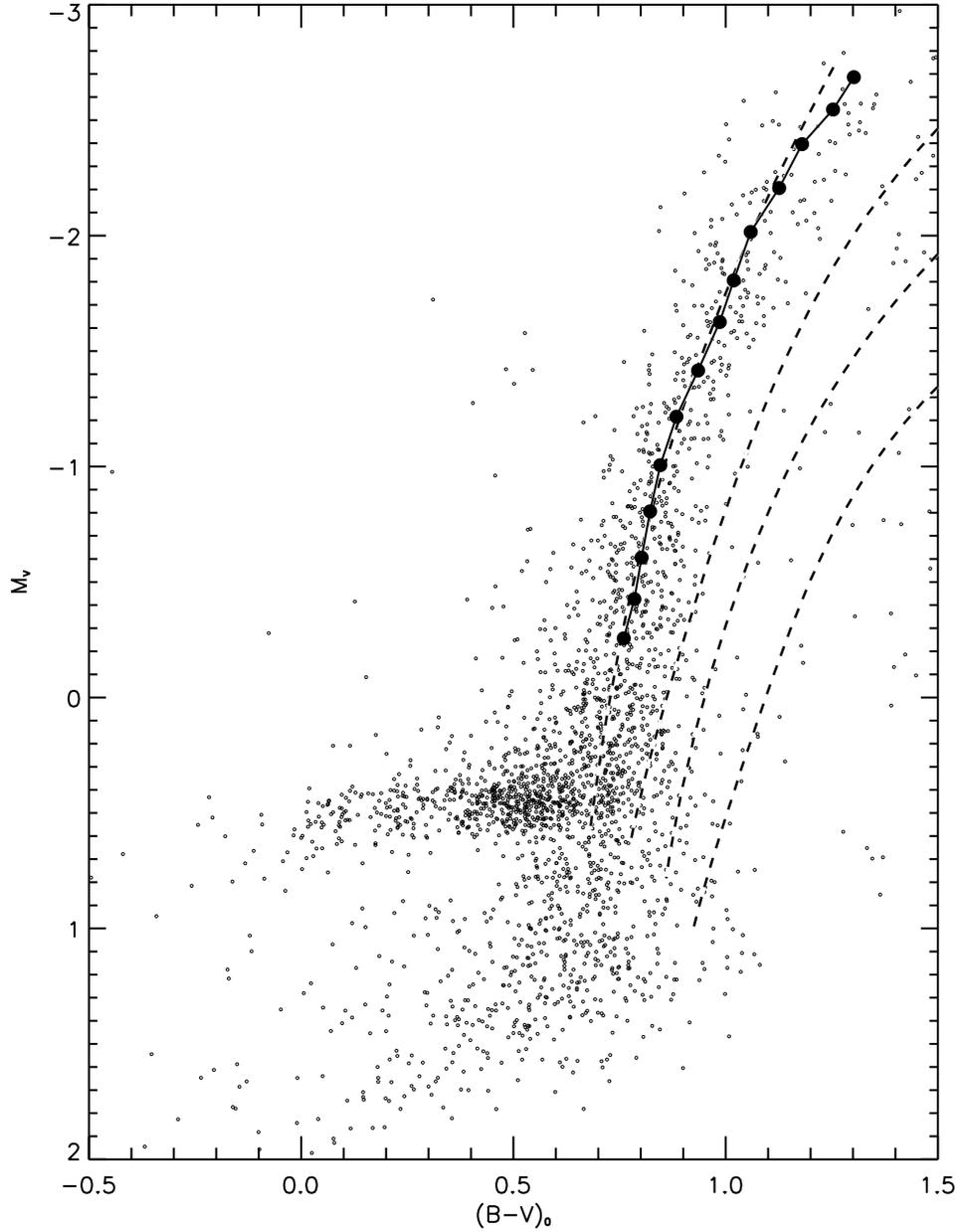}
\caption{The distance and reddening corrected CMD of And V along with our derived
mean red giant branch sequence (filled circles connected by a solid line).  
Overplotted as dashed lines are the Galactic globular 
cluster fiducials of (left to right) M15, NGC 6752, NGC 1851, and 47 Tucanae from Sarajedini \&
Layden (1997) representing metallicities of --2.17, --1.54, --1.29, and --0.71, respectively, on
the Zinn \& West (1984) scale.}
\end{figure}

\begin{figure}
%Fig. 5
\epsscale{0.9}
\plotone{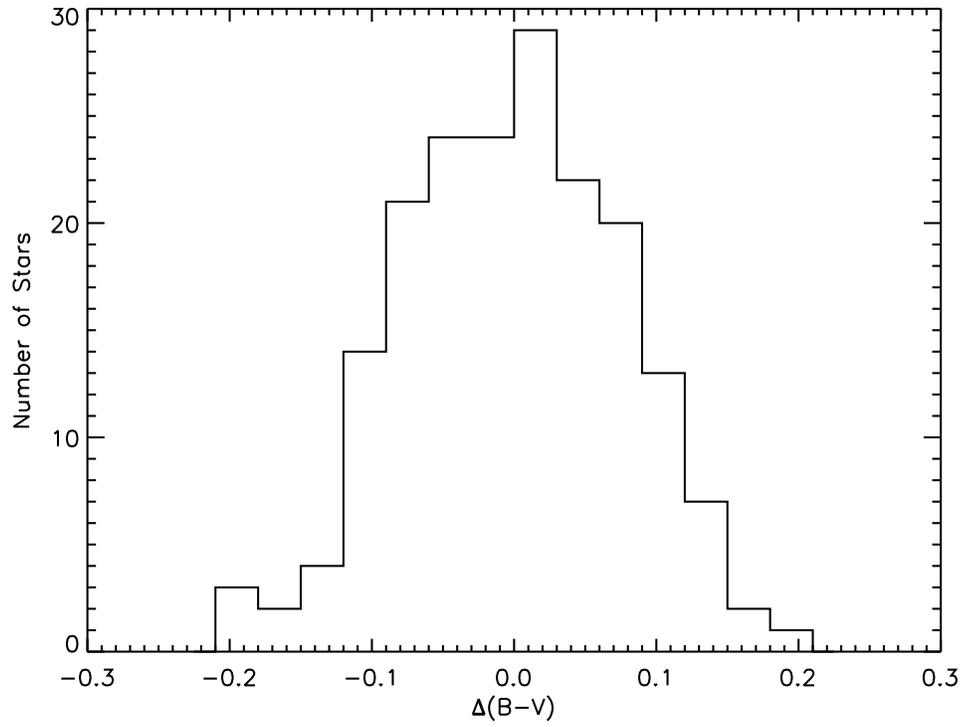}
\caption{The color histogram of red giant stars located in the range $23<V<24$. The color
differences are measured relative to the mean RGB sequence illustrated in Fig. 4.}
\end{figure}

\begin{figure}
%Fig. 6
\epsscale{0.9}
\plotone{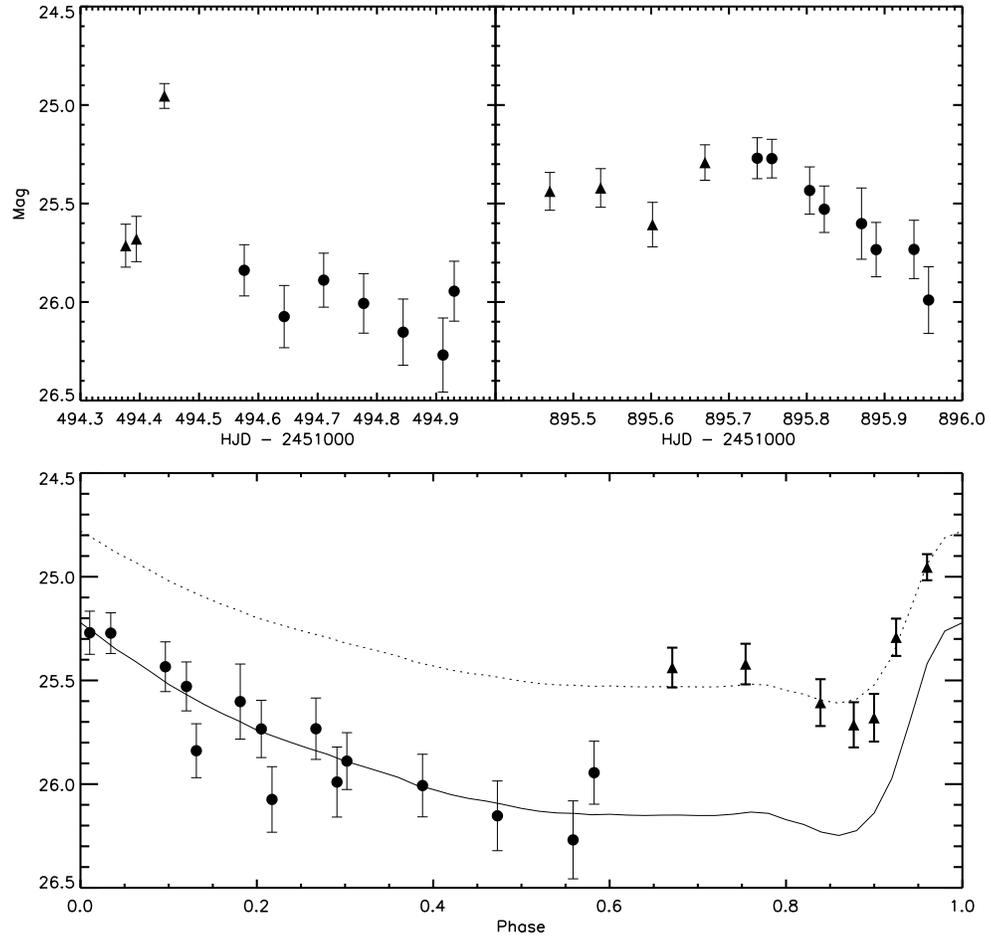}
\caption{The raw and folded light curves of high confidence variable 8.  The upper left and upper right plots are the raw light curves for the two observing windows.  The bottom plot contains the folded light curves and fitted templates.}
\end{figure}

\end{document}